\documentclass[5p,times,twocolumn]{elsarticle}
\usepackage{tikz}
\usetikzlibrary{shapes.geometric, arrows}
\usepackage{graphicx}
\usepackage{txfonts}
\tikzstyle{startstop} = [rectangle, rounded corners, minimum width=3cm, minimum height=1cm,text centered, draw=black, fill=red!10]
\tikzstyle{io} = [trapezium, trapezium left angle=70, trapezium right angle=110, minimum width=2cm, minimum height=1cm, text centered,text width=2.5cm, draw=black, fill=blue!10]
\tikzstyle{process} = [rectangle, minimum width=2.5cm, minimum height=1cm, text centered, text width=2cm, draw=black, fill=orange!10]
\tikzstyle{decision} = [diamond, minimum width=2cm, minimum height=1cm, text centered, text width=2cm, draw=black, fill=green!10]

\tikzstyle{arrow} = [thick,->,>=stealth]
\usepackage{comment}
\usepackage{vmargin}
\usepackage{makecell}
\usepackage{diagbox}
\usepackage{blindtext}
\usepackage{subfig}
\usepackage{natbib}
\setcitestyle{authoryear,aysep={,},citesep={;},open={},close={}}

\usepackage{amssymb}

\usepackage[figuresright]{rotating}

\begin{document}

\begin{frontmatter}

\title{Stellar Spectra Models Classification and Parameter Estimation Using Machine Learning Algorithms}

\author[1]{Miguel Flores R.}
\author[1,2]{Luis J. Corral}
\author[3]{Celia R. Fierro-Santillán}
\address[1]{Centro Universitario de Ciencias Economico Administrativas, Universidad de   
            Guadalajara, 45100 Zapopan, Jal., México.}
\address[2]{Instituto de Astronomía y Meteorología (IAM), Universidad de Guadalajara, 44130 Guadalajara, Jal., México.}
\address[3]{Departamento de Física, Instituto Nacional de Investigaciones Nucleares (ININ), 52740 Ocoyoacac, Estado de México, Mexico}

\begin{abstract}
The growth of sky surveys and the large amount of stellar spectra in the current databases, has generated the necessity of developing new methods to estimate stellar parameters, a fundamental task on stellar research. In this work we present a comparison of different machine learning algorithms, used for the classification of stellar synthetic spectra and the estimation of fundamental stellar parameters including $T_{eff}(K)$, $log(L/L_o)$, $log$ $g$, $M/M_o$, and $V_{rot}$. For both tasks, we established a group of supervised learning models, and propose a database of measurements with the same structure to train the algorithms. These data include equivalent-width measurements over noisy synthetic spectra in order to replicate the natural noise on a real observed spectrum. Different levels of signal to noise ratio are considered for this analysis.

\end{abstract}


\end{frontmatter}

\section{Introduction}
The automatic fit of synthetic spectra to get the right parameters of an observed signal can be a very complex task. Achieving a high accuracy will depend on the number and type of parameters used to create the models, the number of points the models generate for every spectrum, and from the other side, the quality of the observed signal. Over the last years, different types of algorithms were proposed to reach this objective, from direct analysis and discrimination on measurements over both, synthetic and observed spectrum (\citep{Worley_2012, Recio-Blanco_2006, Fierro_2018, Sander_2015, Fierro_2021,Hamann_2003}), to more modern algorithms including artificial neural networks (ANNs) (\citep{Teimoorinia_2012, Dafonte_2016, Minglei_2017, Li_2017, Navarro_2012, Villavicencio_2020, Sharma_2019, Bu_2020, Sharma_2020,Snider_2001, Bin_2020}). This type of modern algorithms has shown very good results in different projects that seek to estimate chemical composition, specific stellar parameters, etc. (\citep{Navarro_2012, Villavicencio_2020, Sharma_2019, Bu_2020}). Because of the complexity to build and calibrate this type of algorithms, the number of parameters that the systems predict differs from project to project. In this context, we propose as the first step of this project, the implementation of different types of machine learning algorithms such as random forest classifier and regressor (depends on the task), k-nearest neighbors, Naive Bayes classifier, and artificial neural networks. The comparison of the overall performance between an artificial neural networks and a none network structure algorithm can be a good starting point to design the best system, and satisfy the requirements of a specific project. In this comparison context, the best algorithm for the classification task will be the one that predicts the closest model in the parameters space, and for the regression task, we consider as the best algorithm, the one with the less error in the value prediction of every one of the parameters ($T_{eff}(K)$, $log(L/L_o)$, $log$ $g$, $M/M_o$, and $V_{rot}$), and the average error in the five parameters prediction. 

In past projects (\citep{Recio-Blanco_2006, Koleva_2009, Ting_2016, Fierro_2018}), the automated fit of synthetic stellar spectra has shown the big improvement that represents reducing the computational time for processing the big amount of astronomical data observed in the different telescopes and projects over the world. But since the first projects that tried to automate some of the processes to accurately fit a stellar spectrum, developing a reliable system that can fit any type of spectrum or most of the spectra from different types of star and with different levels of noise available in some databases, seems to be a very difficult task, and some algorithms simply can not process that data. In \citep{Worley_2012}, is mentioned the issue when an algorithm can not handle some types of spectra or data with not enough quality. Because of this, we propose a new system based on the knowledge that past approaches give to us, using modern computer tools on the machine learning field like the artificial neural networks in their different types of structures.

The present work studies the problem of finding the best model that fits an observed spectrum. This type of task can be defined in two different approaches. The first approach can be seen as a classification problem, where an hyper-variable that labels a single model, represent a sequence of specific parameters, then a machine learning system will try to find this label based on the inputs. The second approach is the estimation of the minimal necessary parameters that define the best model.The traditional approach is to compare spectra using the chi-square. This work takes a different perspective and establishes a new solution based on the equivalent widths of the spectral lines, which reduces the number of microprocessor operations and avoids dealing with the resolution of the spectra. Under this approach, we do not fit a spectral range, but the equivalent widths of a set of lines.

Also in this approach, we can take two routes, develop one system to estimate a single parameter (many to one problem), and develop a system to estimate multiple parameters at the same time (many to many problems). Additionally, in this paper, we propose a number of parameters needed to train the machine learning systems and reduce the dimensionality of the problem, and with this decision avoid the necessity of processing all spectra points of every model and also the requisite of implement the system in a super computer. Finally, following this philosophy, we are working on the TensorFlow back-end and also using the Keras library as most of the works nowadays to develop all the artificial neural networks, and for the non-network algorithms, we use the scikit-learn library. These libraries are easy implementation open source, which allows users to easily build the system.

In this paper, section 2 describes the structure of the models and measurements over these models to create the input parameters for the systems. Also, we introduce the artificial neural network, random forest, k-nearest neighbors, and Naive Bayes algorithms. Section 3 reports some predictions and analyze the performance of the machine learning models, and in section 4, we take the best algorithms to make and extra test. Finally, our work is summarized in section 5.

%\section{Methods}
\section{Database Collection}

\subsection{The Models}
For this work we use the models of OB type stars developed by \citep{Fierro_2015}, and \citep{Zsargo_2020}, that are been used to fit stellar observed spectra and compared with other stellar spectra models in this task. For this starting project, we used a total of 3734 synthetic spectra models from their database. 

The models cover stars with mass from 9 to 120 $M\textsubscript{\(\odot\)}$, and each synthetic spectrum calculated in the optical (3500–7000 \AA) and composed by 100,001 points with uniform steps of 0.035 \AA.

\begin{table}[!htbp]
%\vspace{-0.1 cm}
\small
\setlength\tabcolsep{1 pt}
\centering
    \begin{tabular}{ll}
    \hline
        Parameters & \hspace{-2cm} Value  \\
        \hline
        $T_{eff}$ &\hspace{-2cm} Effective Temperature$^{a}$  \\
        L&\hspace{-2cm}  Luminosity$^{a}$  \\
        Z &\hspace{-2cm} Solar Metallicity$^{a}$ \\
        \textit{v}$_{\infty}$ &\hspace{-2cm} Terminal Velocity: 2.1 \textit{v}$_{esc}$\\
        $\beta$ &\hspace{-2cm} 0.5, 0.8, 1.1, 1.4, 1.7, 2.1 2.3\\
        F$_{cl}$ &\hspace{-2cm} 0.005, 0.30, 0.60, 1.00\\
        M &\hspace{-2cm} Stellar Mass$^{a}$ \\
        R &\hspace{-2cm} Stellar Radius from: M and L  \\
        log g &\hspace{-2cm} Surface Gravity from: M and L\\
        \textit{v} sin i &\hspace{-2cm} Apparent Velocity\\
        $\dot{M}$ &\hspace{-2cm} Mass Loss Rate \\
        \hline
        \makecell{\textbf{Note.} $^{a}$ Atlas of CMFGEN Models\\ (\citep{Fierro_2015}) }\\
    \end{tabular}
    \caption{Stellar Parameters}
    \label{tab:Stellar Parameters}
\end{table}

These models are defined in a 6D space, where each dimension corresponds to a one of the first six parameters showed in Table \ref{tab:Stellar Parameters}, however, one can have all the eleven parameters. The models also include the velocity of the stellar wind at a large distance from the star $v_{\infty}$, the $\beta$ exponent of the velocity law defined in Equation 1, and it controls how the stellar wind is accelerated to reach the terminal velocity. Because the stellar wind is not necessarily homogeneous, it is assume that it contains gas in the form of small clumps or condensations, such that, $F_{cl}$ is the clumping factor, fraction of the total volume occupied by the gas clumps, while the space between them is assumed to be a vacuum.

\begin{equation}
    v(r) = v_{\infty}\left(1-\frac{r}{R_{\star}}\right)^{\beta}.
\end{equation}

The rotational speeds of the models correspond to the evolutionary data of \citep{ekstrom_2012}, starting at the ZAMS with equatorial rotation speed $V_{eq}$ = 0.4  $V_{crit}$, where $V_{crit}$ is the critical velocity, which decreases as the star loses angular momentum due to loss of mass or magnetic braking.

%Low values of $\beta$ indicate fast wind acceleration and high values indicate lower accelerations

\begin{figure}[!ht]
\hspace*{-0.5 cm}
    \centering
    \includegraphics[scale=0.37]{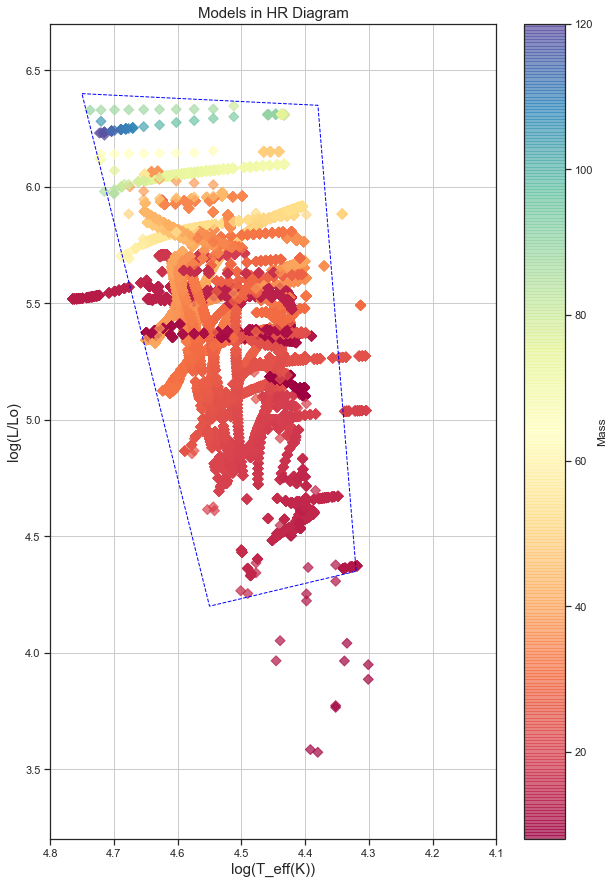}
    \caption{Database of models in the HR diagram group by the mass of the star, where the dot polygon is the area with more density of models.}
    \label{fig: Models Database}
\end{figure}

\subsection{Dimensionality Reduction}
We proposed a method to reduce the amount of data that gets into the machine learning systems as the input features. Starting on the system proposed in \citep{Fierro_2018}, we expand the number of absorption lines measurements, in order to have information of lines in the red part of the visual spectrum, and also lines of heavier elements such as SiIII, SiIV, etc. The lines of SiII (4128) and SiIII (4552), which are useful to restrict the Temperature of B stars, were excluded because most of the spectra are faint, with an equivalent width of thousandths or minor. Additionally, the SiII (4128) line has several neighboring lines, which increases the uncertainty of its equivalent width. 

We choose 34 absorption lines that we think can be the minimal number to get high accuracy for both approaches. These 34 lines are shown in Table \ref{tab: Lines}.  For these lines, we check the possible neighbor lines that can be blended. The principal line is chosen as the closest Ion to the centroid of the absorption line in the spectrum.

\begin{table}[!ht]
\small
\setlength\tabcolsep{1pt}
    \centering
    \begin{tabular}{c|c|c|c}
    \makecell{Wavelength \\ Centroid (\AA)} & \makecell{Principal \\ Line} & Ions & \makecell{Total \\ Ions}\\
    \hline
        3750 & H12 & NIV, H12, NIII, OIII, OIII & 5 \\
        3770 & H11 & H11, SiIV, OIII, NeII, OII & 5 \\
        3798 & H10 & NeII, OIII, SiII, H10, OII & 5 \\
        3820 & HeI & HeI & 1 \\
        3835 & H9 & OII H9 & 2 \\
        3889 & H8 & OII, CII, H8 & 3 \\
        3926 & HeI & NIV, H12, NIII, OIII, OIII & 2 \\
        3934 & CaII & CaII & 1 \\
        3970 & H$\epsilon$ & HeI, HeII, CaII, H$\epsilon$ & 4 \\
        4026 & HeI & HeI & 1 \\
        4058 & NIV & NIV & 1\\
        4088 & SiIV & SiIV & 1 \\
        4101 & H$\delta$ & CaII, H$\delta$ & 2 \\
        4143 & HeI & HeI & 1 \\
        4200 & HeII & HeII & 1 \\
        4212 & SiIV & SiIV & 1 \\
        4340 & H$\gamma$ & OII, OII, H$\gamma$ & 3 \\
        4388 & HeI & HeI & 1 \\
        4471 & HeI & HeI & 1 \\
        4481 & MgII & MgII & 1\\
        4541 & HeII & HeII & 1 \\
        4813 & SiIII & SiIII & 1 \\
        4861 & H$\beta$ & HeII, H$\beta$ & 2 \\
        4943 & OII & OII, OII & 2\\
        4987 & NII & NII & 1 \\
        5005 & NII & NII, NII, NII, NII, NII, NII, & 6\\
        5016 & HeI & HeI & 1\\
        5048 & HeI & NII, HeI & 2\\ 
        5801 & CIV & CIV & 1 \\
        5812 & CIV & CIV & 1 \\
        6381 & NIV & NIV & 1\\
        6527 & HeII & HeII & 1 \\
        6563 & H$\alpha$ & HeII, H$\alpha$ & 2 \\
        6683 & HeII & HeII & 1 \\
    \end{tabular}
    \caption{Total lines needed as an input for the ANNs in the new approach}
    \label{tab: Lines}
\end{table}

For the election of the lines, we also considered the lines that are easy to measure in most of the synthetic spectra, like isolated lines, lines where it is easy to establish the start and the end of the line with respect to the normalized continuum. In some cases, these problems can not be avoided so we also proposed a better way to consistently measure the equivalent width of the lines, using the theoretical centroid of the lines we take $centroid + \alpha$ as the right limit and $centroid - \alpha$ as the left limit, and measure the equivalent-width between this limits. Where $\alpha$ is unique for every line, and was established as a distance from the centroid of the line to near the continuum trying to avoid the next absorption or emission line. This type of alternative measure to the classical equivalent width proposed in \citep{Worthey_1994} and used recently by \citep{Navarro_2012}, is an efficient way to determine changes on absorption lines.

\section{Methods}
\subsection{Artificial Neural Networks}
Since the results obtained in different projects using artificial neural networks (ANNs) (\citep{Navarro_2012, Villavicencio_2020, Sharma_2019, Bu_2020, Sharma_2020, Snider_2001}), we proposed a system based on the recurrent network structure, and following the idea of reduce the necessity of high computational resources, we explore the simple network structures and non-network structures needed to complete the tasks.

The election of this type of network structure is based on the type of input data, which is a numerical vector with an specific sequence for each spectrum, and the values will change in relation with the noise in the spectrum. Additional to this, the system needs to have is the capacity to distinguish between the nearby models. One classic solution to the systems with very similar possible outputs is the connection between the neurons and also the feedback connections.

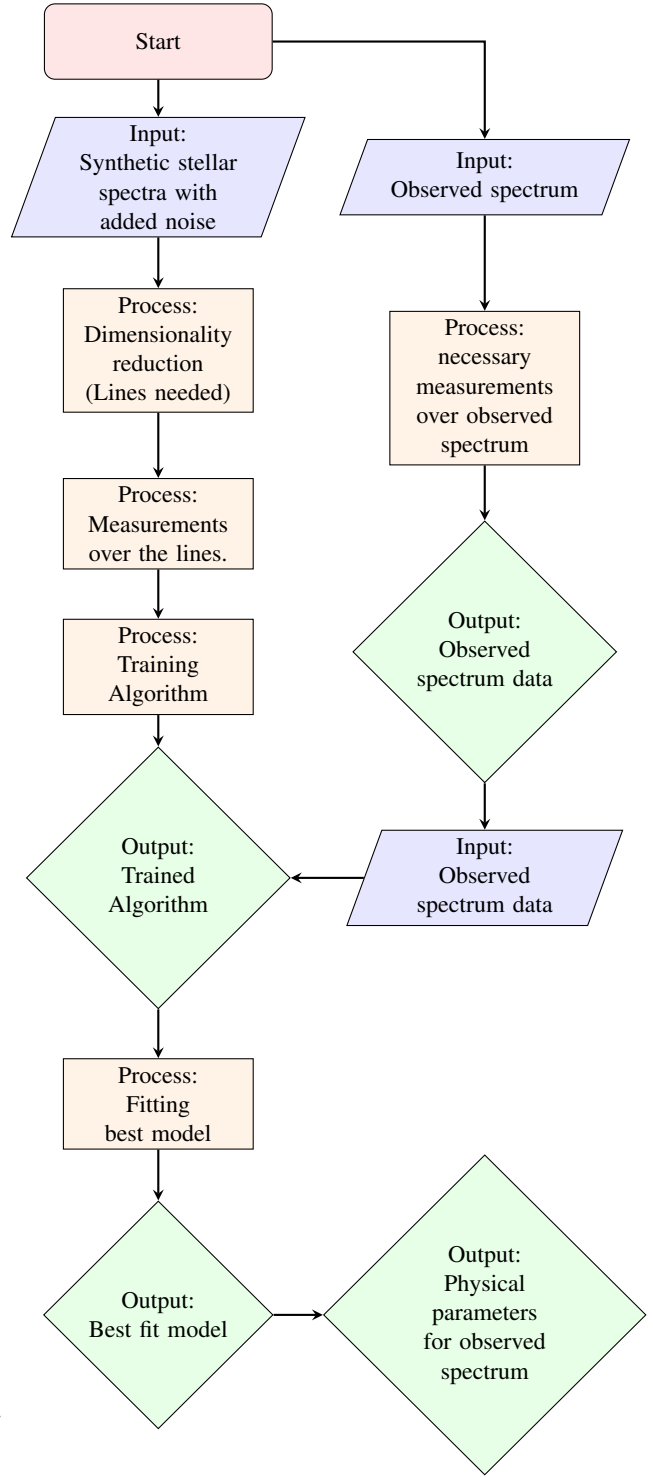
\begin{figure}[!ht]
\hspace*{-0.6 cm}
%\hspace{0.2cm}
\small
\setlength\tabcolsep{0.5pt}
%\centering
    \begin{tikzpicture}[node distance=1.8cm]
    \node (start) [startstop] {Start};
    \node (in1) [io, below of=start]{Input:\\ Synthetic stellar spectra with added noise};
    \node (in1a) [io, right of=in1, xshift=2.5cm]{Input:\\ Observed spectrum};
    \node (pro1a) [process, below of=in1, yshift=-0.5cm] {Process: \\ Dimensionality reduction (Lines needed)};
    \node (pro1) [process, below of=pro1a, yshift=-0.5cm] {Process: \\ Measurements over the lines.};
    \node (pro1o) [process, right of=pro1a, xshift=2.5cm, yshift=-0.5cm] {Process:\\ necessary measurements over observed spectrum};
    \node (dec1o) [decision, below of=pro1o, yshift=-1.7cm] {Output: \\ Observed spectrum data};
    
    \node (pro2) [process, below of=pro1, yshift=-.1cm] {Process: \\ Training Algorithm};
    \node (dec1) [decision, below of=pro2, yshift=-1cm] {Output:\\ Trained Algorithm};
    \node (in2) [io, right of=dec1, xshift=2.5cm] {Input:\\ Observed spectrum data};
    \node (pro3) [process, below of=dec1, yshift=-1.2cm] {Process: \\Fitting best model};
    \node (dec2) [decision, below of=pro3, yshift=-1cm] {Output:\\ Best fit model};
    \node (dec3) [decision, right of=dec2, xshift=2.5cm] {Output:\\Physical parameters for observed spectrum};
    
    \draw [arrow] (start) -- (in1);
    \draw [arrow] (start) -| (in1a);
    \draw [arrow] (in1a) -- (pro1o);
    \draw [arrow] (pro1o) -- (dec1o);
    \draw [arrow] (dec1o) -- (in2);
    
    \draw [arrow] (in1) -- (pro1a);
    \draw [arrow] (pro1a) -- (pro1);
    \draw [arrow] (pro1) -- (pro2);
    \draw [arrow] (pro2) -- (dec1);
    \draw [arrow] (in2) -- (dec1);
    \draw [arrow] (dec1) -- (pro3);
    \draw [arrow] (pro3) -- (dec2);
    \draw [arrow] (dec2) -- (dec3);
    \end{tikzpicture}
    \caption{Steps followed by the system, where the algorithm can be an artificial neural network or another machine learning algorithm.}
    \label{fig: Flux Diagram}
\end{figure}

The full system that we propose follows the next steps: First, we modify the models adding different levels of white noise in order to obtain spectra with high noise level (20 S/N) and low noise (115 S/N), this part of the system has the purpose of providing noisy spectra to the machine learning algorithms, and generate the capacity to deal with future observed spectra which naturally have noise. The next step is reducing the dimensionality of the data, this is as we mentioned, measure specific lines as a similar way to an equivalent width measurement. The final step is training the machine learning algorithms, as most of the current projects, we randomly take $80\%$ of the total models to make the training process and the other $20\%$ to do the test process. On Figure \ref{fig: Flux Diagram} we show step by step the process of the system.

For the training process, we use a total of 25718 synthetic spectra, 3674 base models with 7 different levels of signal to noise ratio (S/N) plus the clean spectrum. Also, in order to make a similar fitting process that any user will do with observed spectra, we established an "extra-test" database with 420 models that consist of 60 base models that we randomly extract from the full database, this models also with 7 different levels of S/N.
This part is important because we want to be sure that the system can deal with different synthetic noise levels and improved the system before it is in contact with an observed spectrum that does not have a controlled S/N.

For the classification task, after the system finds the best model ID that can fit the input spectrum, the next step is to extract the physical parameters (showed on Table \ref{tab:Stellar Parameters}) in the model's database and assign them to the observed spectrum. 

The artificial neural network structures that we proposed, are the classical recurrent neural network (RNN) because the connections of the nodes are a useful feature to deal with multiple input tasks, and the Long Short-Term Memory (LSTM) network, a most robust recurrent neural network. The election of the LSTM network, is because its capacity to deal with sequence numerical problems where the problem is a many features to one output variable.

\subsection{Another Machine Learning Algorithms Approach}
In order to design the best system for the classification and regression problems, we tried different machine learning algorithms. For classifications task: Random Forest Classifier (RFC), Decision Tree Classifier (DTC),  K-Nearest Neighbors (KNN), Gaussian Naive Bayes (GNB). For the regression task we use the Random Forest Regressor (RFR) as the model to compare with the ANNs.

The election of this group of algorithms has the only purpose of analyze the different perspectives that can solve the classification and regression tasks. The Decision Tree Classifier is a system that split in multiple evaluation nodes the features in order to established if an specific group of features belongs to a class, the route over the tree with the best evaluation based on the selected metrics will establish the class, and the Random Forest is an expansion of the Decision Tree composed by a group of Trees providing more nodes to measure over the different features and output classes, with the capacity of split randomly the data on the number of trees and get different evaluations of the group of features that belongs to each class. On the other hand, the GNB follows the Bayes theorem and calculates conditional probabilities assuming the independence of the assumptions that a feature is a member of a specific class. And the KNN measures the distance over the groups of features with the goal of assigning every group of features to one of the $K$ classes. Finally, since the decision tree and consistently the random forest algorithm, can perform a regression task, we chose the random forest as a comparison algorithm, where the output will be the average of the results of each tree in the forest.

These types of algorithms based on mathematical methods like the minimal distance between the parameters that define every output variable and the closest one (KNN), or evaluate the values of the features to estimate the best group of features that characterize the output variable (DTC, RFC), can be a good comparison with the more complex models like the artificial neural networks. These algorithms have shown good results in different projects including the stellar spectra analysis and classification (\citep{Hinners_2018, Sharma_2020, Li_2019}). Moreover, it has been shown that different machine learning algorithms with a non-neural network structure can have similar results as the ANNs depending on the amount and type of data, and also, having the advantage of an easy developing process and less computational time of training. Knowing that not every user will have the same data and necessities this comparison will give a clear idea of the pros and cons of different algorithms doing the same task of analyzing stellar models.

\section{Results and Discussion}

The following section shows the results obtained by the different systems for the classification and regression approaches. Because every system follows the same workflow, with the same input variables but different output, the step by step system diagram showed in Figure \ref{fig: Flux Diagram} can be used for both problems. The user only needs to change the algorithm for every case and the output variables. In the case of the regression task, this workflow needs to be replicated for every unique parameter.

Because every model is built using six physical parameters and there is a specific valid combinations of these parameters, we can take a model-based target as a “hyperparameter” output as we mentioned before.

Another logical approach can be using the six physical parameters as an output, but can be two main problems with this approach: The first one is that because some of the parameters have numerical accuracy of one or two decimal places, and the ANN naturally predict real numbers, and for some physical parameters like the mass where can be models with very close values, the predictions can be equally close to two different values, and predict integers will be a classification task.
The second one is that some of the combinations of the output values of the physical parameters could not exist in the database models. In this part, one can think that if the ANN works well, the output can be a new proposal of parameters for a new model but in this case, we need another section on the system to validate that output, and this is a new complete approach where it is not only fit a model, is also create new models using the predictions of the networks taking the six physical parameters and expand to the other physical parameters needed. 

For the first (and second approach) we use 2 ANN types, as we said above the first one is a simple Recurrent Neural Network structure and  the second one,  an LSTM Network. Also for the LSTM approach, we create two versions, version I is a network with less layers than the version II, this because we want to check the possible number of layers needed in the networks to get similar results, where the network with more layers will need more computational resources and one important part of this work is try to avoid very complex network structures.

On Figures \ref{fig: RFC}, \ref{fig: RNNC}, \ref{fig: HR RFR}, and \ref{fig: HR RNNR} we show some predictions in the HR diagram with the density dotted polygon as Figure \ref{fig: Models Database}. Because classification can predict only existing values and the regression task can predict values that are not in the database, will be useful to see if the algorithms in the classification task predicts models mostly on the area with more density, and for regression task, if the algorithms also predict most of the value in the area with more density of models in the HR Diagram.

\subsection{Classification}
On Tables \ref{tab:ML Algorithms Class} and \ref{tab: RNN Class} we show the evaluation for different machine learning algorithms and neural networks for the classification task using the models identifier as classifier classes and the 34 lines measurements as the input features. 

\begin{figure}[!ht]
    \centering
    \includegraphics[scale=0.37]{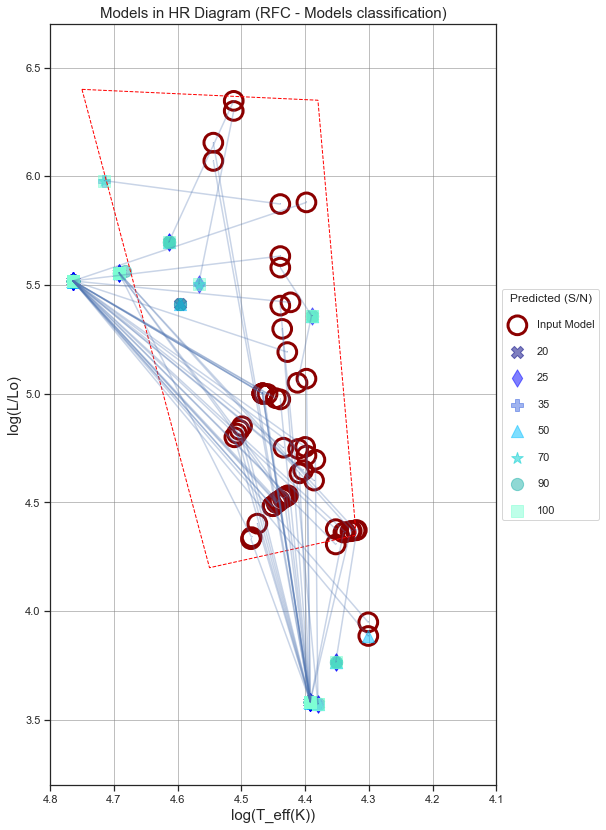}
    \caption{Models classification with Random Forest Classifier in the HR Diagram where the lines represent the distance between the real model and the predicted one.}
    \label{fig: RFC}
\end{figure}

\begin{figure}[!ht]
\hspace{-0.78cm}
    \centering
    \includegraphics[scale=0.37]{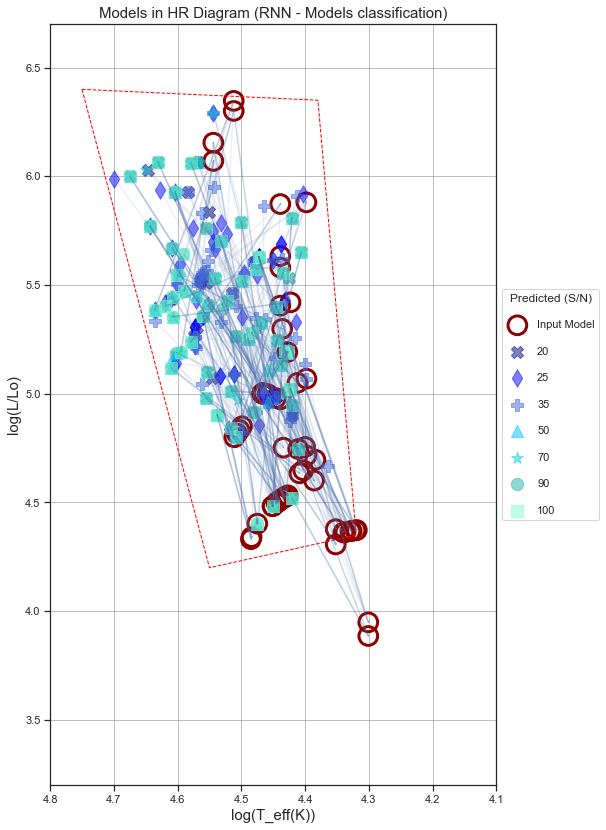}
    \caption{Models classification with Multi-layer Recurrent Neural Networks in the HR Diagram where the lines represent the distance between the real model and the predicted one.}
    \label{fig: RNNC}
\end{figure}

In order to validate and have a metric of the performance of the algorithms, we measure the accuracy of the algorithms for the train and test process, and for the non-network algorithms, we also use the k-cross validations with a $k$=34. For the networks, we measure the train and validation loss in order to avoid over fitting and know how good the neural network is learning, on Table \ref{tab: RNN Class} we show the results obtained at the end of the learning journey, where the loss curves ideally need to follow the same decreasing form and be as close as possible in the final step. 
On Figures \ref{fig: RFC} and \ref{fig: RNNC} we show a comparison between the best ANNs results and the Random Forest Classifier that was the best ML algorithm with non network structure. The results of the additional networks and the DTC, KNN, and GNB, are not shown because we take the models with the best evaluation metrics.

In these two figures we can see as circles the real parameters for the "extra test" that we are doing with every algorithm, and the filled figures the parameters for the predicted models. On the right side, the Recurrent Neural Network, seems to have better results, since the predicted models are closest to the real ones, also the network can predict more variety of models, while the Random Forest apparently only predicts the most distinctive ones. 
One similar behavior is that both models have problems with models that represent stars with more than 70 solar masses, but the lack in the Random Forest is quite bigger, going from 19 solar masses to 85 without any prediction in between.

For the classification task, we only show results for the two cases of simple recurrent network, simple and multi-layer, even though we develop other network structures as different LSTM networks. 
Setup the LSTM networks to get similar results as the recurrent networks have been a complicated task, but since this type of network has been used to classify systems where the input are sequences of complex features, we keep exploring new structures for future work. Even change the shape of the input data, like segmentation based on the levels of noise, can modify the performance of the network.

\begin{table}[!ht]
\setlength\tabcolsep{0.8pt}
    \centering
    \begin{tabular}{c|c|c|c|}
        Classifier & \makecell{Training \\Accuracy} & \makecell{Test\\Accuracy} & \makecell{Cross Validation\\Accuracy}\\
        \hline
        Random Forest & 0.98 & 0.85 & 0.86 \\
        Decision Tree & 0.98 & 0.61 & 0.65 \\
        KNN & 0.88 & 0.83 & 0.85 \\
        GNB & 0.94 & 0.81 & 0.84 \\
    \end{tabular}
    \caption{Non network machine learning algorithms performance results for classification task.}
    \label{tab:ML Algorithms Class}
\end{table}

\begin{table}[!ht]
\setlength\tabcolsep{1pt}
    \centering
    \begin{tabular}{c|c|c|c|c|}
        ANNs & \makecell{Training \\Accuracy} & \makecell{Test\\Accuracy} & \makecell{Train\\Loss} & \makecell{Validation\\Loss}\\
        \hline
        Multilayer & 0.91 & 0.79 & 0.721 & 0.5642\\
        Simple layer & 0.86 & 0.78 & 1.0705 & 1.5150\\
    \end{tabular}
    \caption{Recurrent Neural Networks performance results for classification task.}
    \label{tab: RNN Class}
\end{table}

\begin{table}[!ht]
%\small
\setlength\tabcolsep{5pt}
    \centering
    \begin{tabular}{c|c|c||c}
    Parameter & Real & \makecell{Predicted\\ (S/N=35)} & \makecell{Error\\ ($avg=0.017$)}\\
         \hline
         $T_{eff}(K)$ & 27500 & 27540 & 0.002 \\
         $log(L/Lo)$ & 5.580 & 5.642 & 0.011\\
         $log$ $g$ & 3.135 & 3.070 & 0.021 \\
         $M/M\textsubscript{\(\odot\)}$ & 37 & 36 & 0.027  \\
         $V_{rot}(kms^{-1})$ & 168 & 164 & 0.024 \\ 
        \hline
        \hline
        Parameter & Real & \makecell{Predicted\\ (S/N=25)} & \makecell{Error\\ ($avg=0.020$)}\\
         \hline
         $T_{eff}(K)$ & 22500 & 21890  & 0.027 \\
         $log(L/Lo)$ & 4.307 & 4.362 & 0.013\\
         $log$ $g$ & 3.574 & 3.466 & 0.030 \\
         $M/M\textsubscript{\(\odot\)}$ & 12 & 12 & 0.0  \\
         $V_{rot}(kms^{-1})$ & 158 & 153 & 0.032 \\ 
        \hline
        \hline
        Parameter & Real & \makecell{Predicted\\ (S/N=20)} & \makecell{Error\\ ($avg=0.036$)}\\
         \hline
         $T_{eff}(K)$ & 21890 & 21890  & 0.0 \\
         $log(L/Lo)$ & 4.362 & 4.362 & 0.0\\
         $log$ $g$ & 3.466 & 3.466 & 0.0 \\
         $M/M\textsubscript{\(\odot\)}$ & 12 & 12 & 0.0  \\
         $V_{rot}(kms^{-1})$ & 187 & 153 & 0.182 \\ 
        \hline
        \hline
        Parameter & Real & \makecell{Predicted\\ (S/N=100)} & \makecell{Error\\ ($avg=0.037$)}\\
         \hline
         $T_{eff}(K)$ & 21220 & 21220  & 0.0 \\
         $log(L/Lo)$ & 4.369 & 4.369 & 0.0\\
         $log$ $g$ & 3.405 & 3.405 & 0.0 \\
         $M/M\textsubscript{\(\odot\)}$ & 12 & 12 & 0.0  \\
         $V_{rot}(kms^{-1})$ & 180 & 147 & 0.183 \\ 
        \hline
        \hline
        Parameter & Real & \makecell{Predicted\\ (S/N=100)} & \makecell{Error\\ ($avg=0.037$)}\\
         \hline
         $T_{eff}(K)$ & 27560 & 27670  & 0.004 \\
         $log(L/Lo)$ & 4.508 & 4.504 & 0.001\\
         $log$ $g$ & 3.815 & 3.826 & 0.003 \\
         $M/M\textsubscript{\(\odot\)}$ & 15 & 15 & 0.0  \\
         $V_{rot}(kms^{-1})$ & 241 & 198 & 0.178 \\ 
        \hline
    \end{tabular}
    \caption{Top 5 models predictions with lower average error by the Multiple Layer RNN algorithm for classification task.}
    \label{tab: Results ML RNN}
\end{table}
On Table \ref{tab: Results ML RNN} we show some results for the classification approach using recurrent neural networks, where the predicted physical parameters are the parameters of the model predicted and the S/N value for this predicted model. This table shows the five predictions with less average error, where the error is calculated using a classic error showed on Equation 2. We can see that the top 4 predictions are the same synthetic spectrum with different levels of noise, and the network predicts one close model for the 4 different levels of noise, from 35 S/N to 90 S/N. This behavior can be seen in multiple models in the 420 models we are using to the extra test, where the network predicts one model very close to the real model without being affected by different noise levels. 

\begin{figure}[!htb]
\centering
\hspace{-0.5cm}
\subfloat[ ]{
  \includegraphics[scale=0.3]{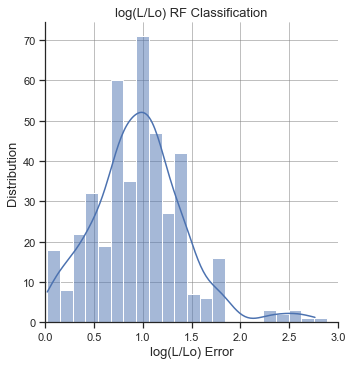}
}
\subfloat[ ]{
  \includegraphics[scale=0.3]{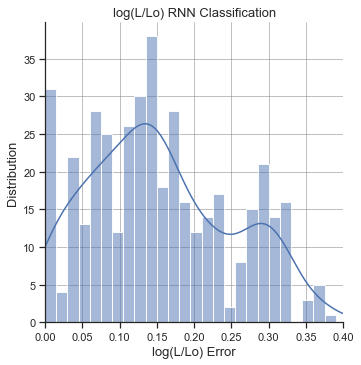}
}
\hspace{-0.5cm}
\subfloat[ ]{
  \includegraphics[scale=0.3]{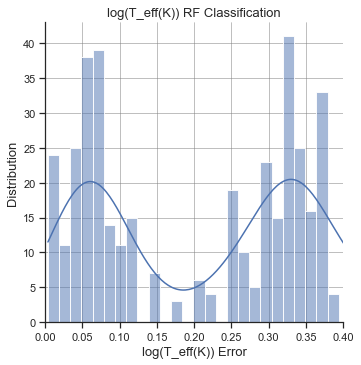}
}
\subfloat[ ]{
  \includegraphics[scale=0.3]{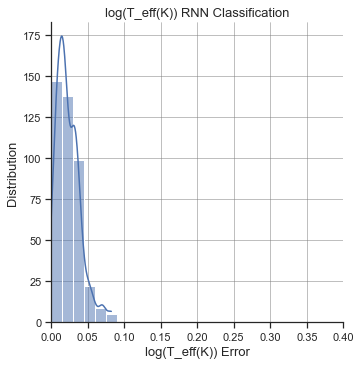}
}
\caption{Error distribution of the predictions for the principal parameters $Log(L/Lo)$ and $log T_eff(K)$ in the classification task. In (a) and (c), the error distribution for the Random Forest Classifier, and in (b) and (d) for the Recurrent Neural Networks}
\label{Classification Error Dens}
\end{figure}

To have another view of the predictions showed on the HR Diagram in Figures \ref{fig: RFC} and \ref{fig: RNNC}, on Figures \ref{Classification Error Dens} and \ref{Regression Error Dens} we show the distribution of the error for the luminosity and temperature parameters using Equation 2, getting an idea of how good or bad was the predictions overall.

\begin{equation}
    Error = \frac{|Input_{P}-Predicted_{P}|}{Input_{P}},
\end{equation}
Where $Real_{P}$ is the input physical parameter and $Predicted_{P}$ is the predicted physical parameter.

\subsection{Regression}
In the context of fitting a stellar spectrum, classification can be a good approach since the algorithm tries to find the closest model, and then, we can establish the physical parameters. According to the results in the last section, the classification task seems to be not as accurate as we expect with the current models and features we are using, such that, we propose a second approach where we design a specific network for some principal parameters, and at the same time, keep the idea of using the random forest regressor to compare the results, because the problem include multiple features in a many to one problem. On Table \ref{tab: RFR Results} we show the results for this algorithm.

\begin{figure}[!ht]
    \centering
    \includegraphics[scale=0.37]{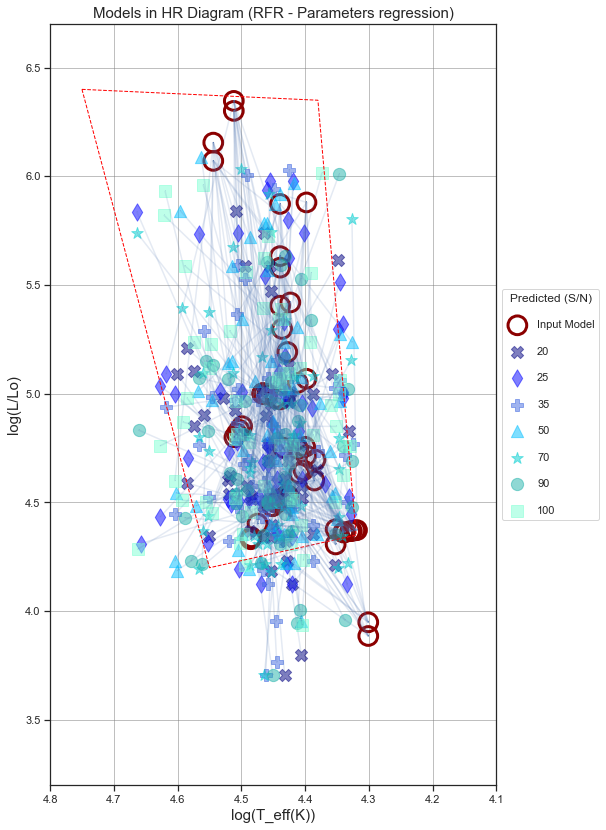}
    \caption{Models parameter regression with Random Forest Regressor in the HR Diagram where the lines represent the distance between the real model and the predicted one.}
    \label{fig: HR RFR}
\end{figure}

\begin{figure}[!ht]
    \centering
    \includegraphics[scale=0.37]{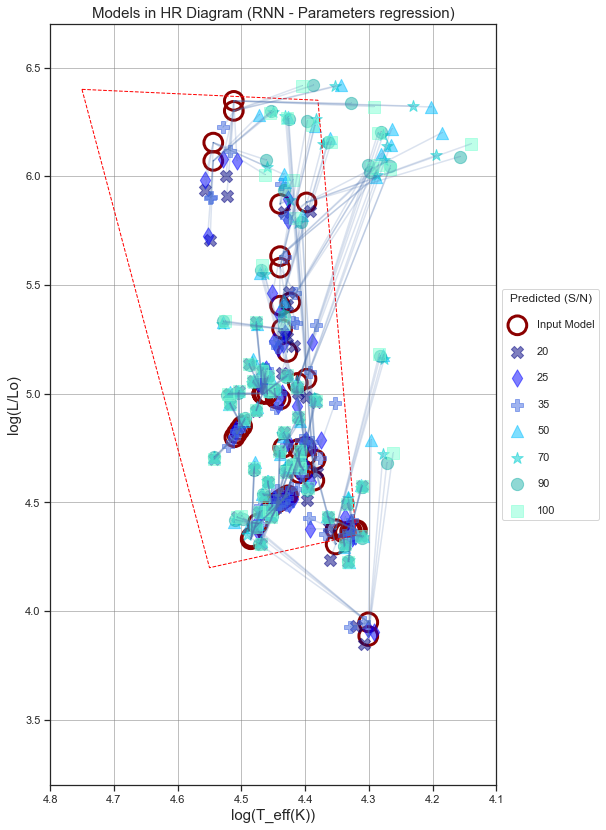}
    \caption{Models parameter regressions with Multi-layer Recurrent Neural Networks in the HR Diagram where the lines represent the distance between the real model and the predicted one.}
    \label{fig: HR RNNR}
\end{figure}

\begin{table}[!ht]
\small
\setlength\tabcolsep{1.2pt}
    \centering
    \begin{tabular}{c|c|c|c}
    \hline
    Parameter &  \makecell{Unique\\Parameters} & Min Value & Max Value\\
         \hline
         $T_{eff}(K)$ & 1255  & 20000 & 58140\\
         $M/M\textsubscript{\(\odot\)}$ & 1853 & 8 & 120\\
         $log(L/Lo)$ & 1103 & 3.573 & 6.347\\
         $log$ $g$ & 1184 & 2.574 & 4.690\\
         $V_{rot}(kms^{-1})$  & 263 & 122 & 473\\ 
        \hline
    \end{tabular}
    \caption{Range of total parameters values in database}
    \label{tab: Total parameters}
\end{table}

\begin{table}[!ht]
\small
\setlength\tabcolsep{1.2pt}
    \centering
    \begin{tabular}{c|c|c|c}
         \makecell{Random Forest\\Regressor}\\
    \hline
    \hline
    \backslashbox{Parameter}{Accuracy} &  Train & Test & Extra Test\\
         \hline
         $T_{eff}(K)$ & 0.97 & 0.97	& 0.77 \\
         $log(L/Lo)$ & 0.98 & 0.97 & 0.83\\
         $log$ $g$ & 0.99 & 0.97 & 0.90 \\
         $M/M\textsubscript{\(\odot\)}$ & 0.99 & 0.98  & 0.41\\
         $V_{rot}(kms^{-1})$  & 0.99 & 0.98 & 0.79 \\ 
        \hline
    \end{tabular}
    \caption{Random Forest Regressor}
    \label{tab: RFR Results}
\end{table}

\begin{table}[!ht]
\small
    \centering
    \begin{tabular}{c|c|c|c}
         \makecell{Bidirectional\\LSTM}\\
    \hline
    \hline
    \backslashbox{Parameter}{Accuracy} &  Train & Test & Extra Test\\
         \hline
         $T_{eff}(K)$ & 0.97 & 0.97	& 0.90 \\
         $log(L/Lo)$ & 0.95 & 0.95 & 0.94\\
         $log$ $g$ & 0.93 & 0.93 & 0.93  \\
         $M/M\textsubscript{\(\odot\)}$ & 0.93 & 0.93  & 0.78\\
         $V_{rot}(kms^{-1})$ & 0.98 & 0.98 & 0.90 \\ 
        \hline
    \end{tabular}
    \caption{Multiple Layer Bidirectional LSTM}
    \label{tab: MLB LSTM}
\end{table}

\begin{table}[!ht]
\small
    \centering
    \begin{tabular}{c|c|c|c}
         \makecell{Bidirectional\\LSTM}\\
    \hline
    \hline
    \backslashbox{Parameter}{Accuracy} &  Train & Test & Extra Test\\
         \hline
         $T_{eff}(K)$ & 0.83 & 0.83	& 0.71 \\
         $log(L/Lo)$ & 0.95 & 0.95 & 0.95\\
         $log$ $g$ & 0.93 & 0.93 & 0.95  \\
         $M/M\textsubscript{\(\odot\)}$ & 0.93 & 0.93  & 0.69\\
         $V_{rot}(kms^{-1})$ & 0.97 & 0.97 & 0.77 \\ 
        \hline
    \end{tabular}
    \caption{Simple Layer Bidirectional LSTM}
    \label{SLB LSTM}
\end{table}

\begin{table}[!ht]
\small
\setlength\tabcolsep{1pt}
    \centering
    \begin{tabular}{c|c|c|c}
         \makecell{RNN}\\
    \hline
    \hline
    \backslashbox{Parameter}{Accuracy} &  Train & Test & Extra Test\\
         \hline
         $T_{eff}(K)$ & 0.98 & 0.98	& 0.88 \\
         $log(L/Lo)$ & 0.95 & 0.95 & 0.93\\
         $log$ $g$ & 0.99 & 0.98 & 0.92  \\
         $M/M\textsubscript{\(\odot\)}$ & 0.95 & 0.94  & 0.79\\
         $V_{rot}(kms^{-1})$ & 0.97 & 0.97 & 0.82 \\ 
        \hline
    \end{tabular}
    \caption{Multiple Layer RNN}
    \label{ML RNN}
\end{table}

As we can see in Table \ref{tab: MLB LSTM}, the second version of the LSTM network that has twice the internal layers has better results in general, but as expected needs more processor time to be trained. Even though the results for the training and test apparently showed good performance for version II of the LSTM structure, when we try to fit the models in the extra test data set, none of the three different ANNs predicts more than 15 spectrums in an exact way. That can mean that these specific networks are not truly trained and we can not make predictions and fit spectra that are not in the dataset that is not in contact with the ANNs. Because of that, right now we are exploring different structures of ANNs and identify new lines that we can measure in order to make bigger discrimination of the different types of models.

 Because in the regression task the algorithms need the physical parameters to learn and predict, the distribution of the parameters that define the total space covered by the database are shown in Table \ref{tab: Total parameters}. Even we use different structures of ANNs, the overall time needed to train the networks are pretty similar from one to another. One reason to this behavior and that is also a intentional decision, was to use similar epochs and batch sizes, in order to get consistent results.

For the regression approach, we show in Tables \ref{tab: Results RF Regressor} and \ref{tab: Results RNN Regressions}, and Figures \ref{fig: HR RFR} and \ref{fig: HR RNNR} some of the results using the network with the best performance overall and the Random Forest Regressor as a comparative non-network algorithm. For the RNN as we expect from the accuracy results, the predicted parameters are very accurate for the top 5 models, but we saw that the best 4 are models with a very low level of noise, meanwhile, the results from the RFR shows different levels of noise for the top 5 with less error. In order to have a different analysis we take the top 5 models predicted with less error with a S/N different to 100, that we show in Table \ref{tab: Results RNN Regression None 100}.

\begin{table}[!ht]
%\small
\setlength\tabcolsep{4pt}
    \centering
    \begin{tabular}{c|c|c||c}
    Parameter & Real & \makecell{Predicted\\ (S/N=35)} & \makecell{Error\\ ($avg=0.039$)}\\
         \hline
         $T_{eff}(K)$ & 26850 & 26668 & 0.007 \\
         $log(L/Lo)$ & 4.530 & 4.575 & 0.010\\
         $log$ $g$ & 3.749 & 3.549 & 0.053 \\
         $M/M\textsubscript{\(\odot\)}$ & 15 & 16 & 0.067  \\
         $V_{rot}(kms^{-1})$ & 232 & 219 & 0.056 \\ 
        \hline
        \hline
        Parameter & Real & \makecell{Predicted\\ (S/N=70)} & \makecell{Error\\ ($avg=0.040$)}\\
         \hline
         $T_{eff}(K)$ & 27190 & 27819  & 0.023 \\
         $log(L/Lo)$ & 4.752 & 4.796 & 0.009\\
         $log$ $g$ & 3.543 & 3.437 & 0.030 \\
         $M/M\textsubscript{\(\odot\)}$ & 15 & 16 & 0.067  \\
         $V_{rot}(kms^{-1})$ & 206 & 191 & 0.073 \\ 
        \hline
        \hline
        Parameter & Real & \makecell{Predicted\\ (S/N=25)} & \makecell{Error\\ ($avg=0.042$)}\\
         \hline
         $T_{eff}(K)$ & 29240 & 26519  & 0.093 \\
         $log(L/Lo)$ & 5.001 & 4.466 & 0.107\\
         $log$ $g$ & 3.546 & 3.585 & 0.011 \\
         $M/M\textsubscript{\(\odot\)}$ & 20 & 20 & 0.0  \\
         $V_{rot}(kms^{-1})$ & 223 & 223 & 0.0 \\ 
        \hline
        \hline
        Parameter & Real & \makecell{Predicted\\ (S/N=35)} & \makecell{Error\\ ($avg=0.048$)}\\
         \hline
         $T_{eff}(K)$ & 25030 & 26142  & 0.044 \\
         $log(L/Lo)$ & 4.716 & 4.223 & 0.105\\
         $log$ $g$ & 3.435 & 3.564 & 0.038 \\
         $M/M\textsubscript{\(\odot\)}$ & 15 & 15 & 0.0  \\
         $V_{rot}(kms^{-1})$ & 194 & 184 & 0.052 \\ 
        \hline
        \hline
        Parameter & Real & \makecell{Predicted\\ (S/N=50)} & \makecell{Error\\ ($avg=0.048$)}\\
         \hline
         $T_{eff}(K)$ & 25660 & 26758  & 0.043 \\
         $log(L/Lo)$ & 4.563 & 4.650 & 0.003\\
         $log$ $g$ & 3.563 & 3.438 & 0.035 \\
         $M/M\textsubscript{\(\odot\)}$ & 15 & 15 & 0.0  \\
         $V_{rot}(kms^{-1})$ & 209 & 176 & 0.158 \\ 
        \hline
    \end{tabular}
    \caption{Top 5 predictions with lower average error by the Random Forest Regressor algorithm.}
    \label{tab: Results RF Regressor}
\end{table}
\begin{table}[!ht]
%\small
\setlength\tabcolsep{4pt}
    \centering
    \begin{tabular}{c|c|c||c}
    Parameter & Real & \makecell{Predicted\\ (S/N=100)} & \makecell{Error\\ ($avg=0.007$)}\\
         \hline
         $T_{eff}(K)$ & 25800 & 25758 & 0.002 \\
         $log(L/Lo)$ & 5.050 & 5.002 & 0.010\\
         $log$ $g$ & 3.277 & 3.360 & 0.025 \\
         $M/M\textsubscript{\(\odot\)}$ & 20 & 20 & 0.0  \\
         $V_{rot}(kms^{-1})$ & 156 & 156 & 0.0 \\ 
        \hline
        \hline
        Parameter & Real & \makecell{Predicted\\ (S/N=100)} & \makecell{Error\\ ($avg=0.013$)}\\
         \hline
         $T_{eff}(K)$ & 20000 & 20309  & 0.015 \\
         $log(L/Lo)$ & 3.886 & 3.847 & 0.010\\
         $log$ $g$ & 3.665 & 3.688 & 0.006 \\
         $M/M\textsubscript{\(\odot\)}$ & 9 & 9 & 0.0  \\
         $V_{rot}(kms^{-1})$ & 156 & 161 & 0.032 \\ 
        \hline
        \hline
        Parameter & Real & \makecell{Predicted\\ (S/N=100)} & \makecell{Error\\ ($avg=0.015$)}\\
         \hline
         $T_{eff}(K)$ & 20000 & 20876  & 0.044 \\
         $log(L/Lo)$ & 3.949 & 3.931 & 0.005\\
         $log$ $g$ & 3.602 & 3.605 & 0.001 \\
         $M/M\textsubscript{\(\odot\)}$ & 9 & 9 & 0.0  \\
         $V_{rot}(kms^{-1})$ & 156 & 160 & 0.026 \\ 
        \hline
        \hline
        Parameter & Real & \makecell{Predicted\\ (S/N=100)} & \makecell{Error\\ ($avg=0.016$)}\\
         \hline
         $T_{eff}(K)$ & 22500 & 22991  & 0.022 \\
         $log(L/Lo)$ & 4.307 & 4.235 & 0.017\\
         $log$ $g$ & 3.574 & 3.512 & 0.017 \\
         $M/M\textsubscript{\(\odot\)}$ & 12 & 12 & 0.0  \\
         $V_{rot}(kms^{-1})$ & 158 & 162 & 0.025 \\ 
        \hline
        \hline
        Parameter & Real & \makecell{Predicted\\ (S/N=25)} & \makecell{Error\\ ($avg=0.017$)}\\
         \hline
         $T_{eff}(K)$ & 22500 & 23370  & 0.039 \\
         $log(L/Lo)$ & 4.307 & 4.356 & 0.011\\
         $log$ $g$ & 3.574 & 3.499 & 0.021 \\
         $M/M\textsubscript{\(\odot\)}$ & 12 & 12 & 0.0  \\
         $V_{rot}(kms^{-1})$ & 158 & 160 & 0.013 \\ 
        \hline
    \end{tabular}
    \caption{Top 5 predictions with lower average error by the Multi-layer RNN algorithm for regression task.}
    \label{tab: Results RNN Regressions}
\end{table}

\begin{table}[!ht]
%\small
\setlength\tabcolsep{4pt}
    \centering
    \begin{tabular}{c|c|c||c}
    Parameter & Real & \makecell{Predicted\\ (S/N=25)} & \makecell{Error\\ ($avg=0.017$)}\\
         \hline
         $T_{eff}(K)$ & 22500 & 23370  & 0.039 \\
         $log(L/Lo)$ & 4.307 & 4.356 & 0.011\\
         $log$ $g$ & 3.574 & 3.499 & 0.021 \\
         $M/M\textsubscript{\(\odot\)}$ & 12 & 12 & 0.0  \\
         $V_{rot}(kms^{-1})$ & 158 & 160 & 0.013 \\ 
        \hline
        \hline
        Parameter & Real & \makecell{Predicted\\ (S/N=20)} & \makecell{Error\\ ($avg=0.018$)}\\
         \hline
         $T_{eff}(K)$ & 26791 & 26985  & 0.007 \\
         $log(L/Lo)$ & 5.192 & 5.279 & 0.017\\
         $log$ $g$ & 3.3005 & 3.303 & 0.001 \\
         $M/M\textsubscript{\(\odot\)}$ & 24 & 23 & 0.042  \\
         $V_{rot}(kms^{-1})$ & 156 & 152 & 0.026 \\ 
        \hline
        \hline
        Parameter & Real & \makecell{Predicted\\ (S/N=20)} & \makecell{Error\\ ($avg=0.021$)}\\
         \hline
         $T_{eff}(K)$ & 27300 & 27754  & 0.017 \\
         $log(L/Lo)$ & 5.298 & 5.212 & 0.016\\
         $log$ $g$ & 3.216 & 3.280 & 0.051 \\
         $M/M\textsubscript{\(\odot\)}$ & 24 & 24 & 0.0  \\
         $V_{rot}(kms^{-1})$ & 156 & 148 & 0.051 \\ 
        \hline
        \hline
        Parameter & Real & \makecell{Predicted\\ (S/N=25)} & \makecell{Error\\ ($avg=0.027$)}\\
         \hline
         $T_{eff}(K)$ & 20000 & 20301  & 0.015 \\
         $log(L/Lo)$ & 3.886 & 3.951 & 0.017\\
         $log$ $g$ & 3.665 & 3.946 & 0.077 \\
         $M/M\textsubscript{\(\odot\)}$ & 9 & 9 & 0.0  \\
         $V_{rot}(kms^{-1})$ & 156 & 160 & 0.026 \\ 
        \hline
        \hline
        Parameter & Real & \makecell{Predicted\\ (S/N=90)} & \makecell{Error\\ ($avg=0.027$)}\\
         \hline
         $T_{eff}(K)$ & 28280 & 28670  & 0.014 \\
         $log(L/Lo)$ & 4.483 & 4.367 & 0.004\\
         $log$ $g$ & 3.886 & 4.105 & 0.056 \\
         $M/M\textsubscript{\(\odot\)}$ & 15 & 15 & 0.0  \\
         $V_{rot}(kms^{-1})$ & 251 & 235 & 0.064 \\ 
        \hline
    \end{tabular}
    \caption{Top 5 predictions with lower average error and S/N $\neq$ 100 by the Multi-layer RNN algorithm.}
    \label{tab: Results RNN Regression None 100}
\end{table}

\begin{figure}[!htb]
\hspace{-0.5cm}
\centering
\subfloat[ ]{
  \includegraphics[scale=0.3]{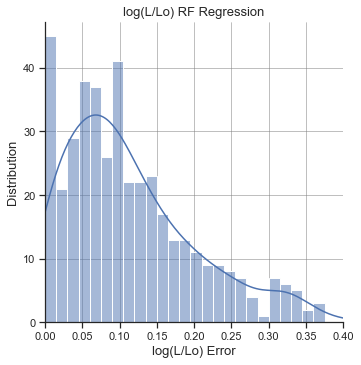}
}
\subfloat[ ]{
  \includegraphics[scale=0.3]{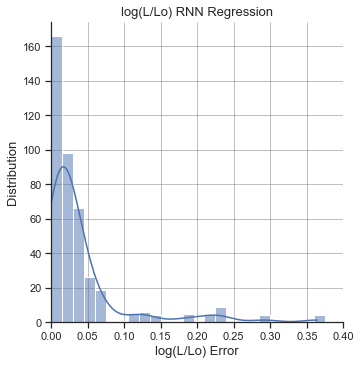}
}
\hspace{-0.5cm}
\subfloat[ ]{
  \includegraphics[scale=0.3]{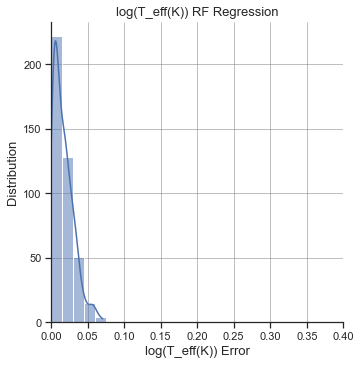}
}
\subfloat[ ]{
  \includegraphics[scale=0.3]{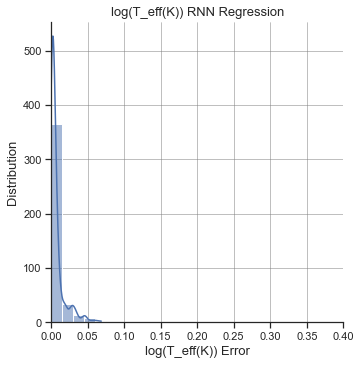}
}
\caption{Error distribution of the predictions for the principal parameters $Log(L/Lo)$ and $log T_eff(K)$ in the regression task. In (a) and (c), the error distribution for the Random Forest Regressor, and in (b) and (d) for the Recurrent Neural Networks}
\label{Regression Error Dens}
\end{figure}

Similar to the classification task, on Figure \ref{Regression Error Dens} we show the error distribution for luminosity and temperature parameters for the regression task that are shown on Figures \ref{fig: HR RFR} and \ref{fig: HR RNNR} in the HR diagram, in order to get a better idea of how good or bad was the results.

As we expect, the ANNs have better results overall for classification and regression task, even we tried to keep the structure of the ANNs as simple as possible, what gives us a good picture to keep working on this approach.
In the classification context, the ANNs have very different behavior, for most of the models, the network predicts a very close model no matter the levels of noise, while the RFC assign different models depending on the level of noise, and tends to repeat more times closest models as we saw on Figures 2 and 3.
For the regression task, as we mentioned before, the random forest regressor demonstrates that a non-neural network algorithm can have similar results as an ANNs in some cases, and also maintains similar responses for some of the parameters like the low accuracy on the estimation of the mass. 
In general, the physical parameters obtained by the ANNs shows less error with a high S/N value (less noise), and for most of the cases the error increase with the level of noise, what is a logical behavior, but some of the best results have the higher noise that is the 20 S/N. After analyzing the results, the ANNs have more problems with some specific models than with the level of noise, since some predictions have similar results for high and low levels of S/N. With the RFR the results are similar, we got high accuracy with high and low levels of S/N, with the exception that this algorithm does not has high accuracy in too many elements with 20 and 25 S/N as the ANNs.

\section{Conclusions}
In the classification task, the Random Forest Classifier takes less time of training compared with the different ANNs we use for this task, the structure of the ideal Forest needs more memory to allocate all the features comparison while the features flows over the trees. Additionally, the memory needed if we add more estimators to the Random Forest grows faster in proportion than the addition of neurons in the ANNs.

In the case of the Random Forest Regressor, the results are good in the training and testing section, but when this algorithm was tested using the \textit{extra test} data sample, the results obtained by the different configurations of the ANNs have better results overall with the same data sample. In this regression task comparison, the individual Random Forest needed to estimate every parameter, needs a fraction of time that any group of ANNs that makes the same task. For example, the best neural network structure to estimate the five physical parameters needs five times the time that the Random Forest Regressor takes to complete the same task.

As we expect for the parameter estimation approach, the ANNs structures have better results, and the difference in training time is not substantial between these different ANNs. For all the different ANNs that we propose, we confirm that the setup of the individual network for every parameter will be different, even that one could propose a very similar structure for every parameter, we can not forget that every problem needs a specific network configuration to be solved, and because the distribution of every the physical parameter in our model database is different one can expect a different learning process in the ANNs. 

Finally, one of the most interesting results is the fact that all tests work well for the ‘not noise’ models (S/N $\sim$ 100), and for the 20 or 25 S/N models, in some specific input spectra models, the noise seems to do not has an effect on the predictions. However, this behavior could be a result of the different representation of stellar models over the physical parameters domain, this can be seen in Figure \ref{fig: Models Database}. One of the next steps of the project will be the expansion of the current database to get a more balance stellar models distribution.

\bibliography{REFS_REVISED.bib}

\end{document}